\documentclass[12pt,amsmath,floats,floatfix,nofootinbib,superscriptaddress]{article}
\usepackage[top=2.5cm,bottom=2.5cm,left=3cm,right=3cm]{geometry}

\usepackage{graphicx} \usepackage{bm} \usepackage{epsfig}
\usepackage{amsmath, amssymb}
\usepackage{wrapfig}
\usepackage{paralist}
   \usepackage{cite}
\usepackage{setspace}
\usepackage{color}
\usepackage{caption}
\usepackage{colordvi}
\usepackage{comment}
\usepackage[export]{adjustbox}
\newcommand{\beq}{\begin{equation}}
\newcommand{\eeq}{\end{equation}}
\newcommand{\bea}{\begin{eqnarray}}
\newcommand{\eea}{\end{eqnarray}}
\newcommand{\bit}{\begin{itemize}\setlength\itemsep{0em}}
\newcommand{\eit}{\end{itemize}}

\newcommand{\nn}{\nonumber}

\usepackage[linktocpage=true]{hyperref}
\hypersetup{
colorlinks=true,
citecolor=blue,
linkcolor=red,
urlcolor=blue,
pdfauthor={},
pdftitle={},
pdfsubject={}
}

\begin{document}
\begin{titlepage}

\begin{center}
{\fontsize{25}{28}\selectfont    The Tune of Love and\\[0.5cm] the {\textit{Nature(ness)}} of Spacetime }
\\[1cm]
{\fontsize{12}{28}\selectfont \large Rafael A. Porto}
\\[0.2cm]
 { \it ICTP South American Institute for Fundamental Research\\ and Instituto de F\'isica Te\'orica - Universidade Estadual Paulista}\\ Rua Dr. Bento Teobaldo Ferraz 271, 01140-070 S\~ao Paulo, SP Brazil\vskip 8pt
 
 \end{center}

\vspace{1cm}
\hrule 
\begin{center}\textbf{Abstract}:\end{center} 
The black hole information paradox is among the most outstanding puzzles~in physics. I argue here there is yet another black hole quandary~which, in light of the recent direct detection of gravitational waves by Advanced LIGO, reveals a new window to probe the nature of spacetime in the forthcoming era of `precision~gravity.' %with GW astronomy.  %with GW observations. 
 \vspace{0.5cm}
\hrule 

\end{titlepage}
%\tableofcontents
\section*{Introduction}
{\it ``One thing is to know sound waves exist, but another thing entirely is to listen to Beethoven's $9^{th}$\,symphony. For the first time we have heard the tune of spacetime.\footnote{\url{https://www.youtube.com/watch?v=TWqhUANNFXw}}} \vskip 12pt

\indent Can nature be unnatural? One would be tempted to answer for the negative, after all this is the rule of thumb when such questions are presented (but see \cite{rule}), and yet the so called `Planck units' for time and space seem completely detached from --not only everyday experience-- experimental accessibility.  Why is the universe so {\it old} and {\it large}? Why is the weak force so {\it strong}? are then some of the questions physicists have tried to address given the minuscule {\it fundamental} units of nature, e.g. \cite{guidice}. Tackling these issues has led to claims about the existence of a multiverse (the landscape) and new spacetime symmetries (supersymmetry), e.g.~\cite{nimaSM,joe}. 
\begin{center}
\includegraphics[width=0.9\textwidth]{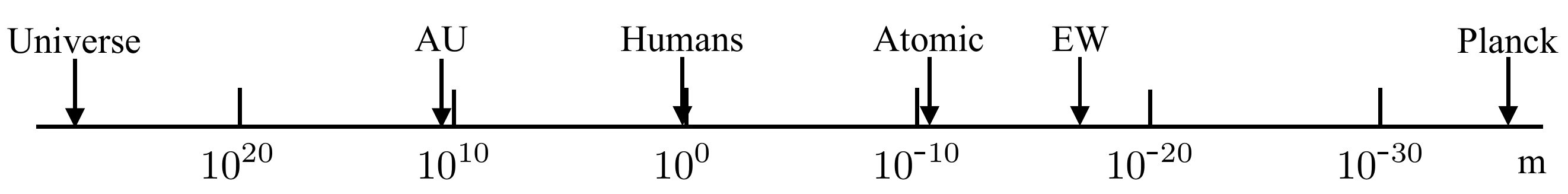}
\end{center}
Black holes --the end product of gravitational collapse-- span a vast range of scales and are expected in almost all sizes, from Planck length ($r_s \sim~\ell_{\rm P}$) to ultra-massive ones ($r_s \gtrsim 10^{48}\, \ell_{\rm P}$) contained in quasars \cite{quasar}.\footnote{The Planck length is given by $\ell_{\rm P}  = \sqrt{\hbar G_N} \sim 10^{-35}$\,m. We use $c=1$ units.}  In fact, they are not devoid of conundrums either. Most notably the `information paradox', or: where did the collapsing {\it stuff} go after Hawking evaporation? This enigma has recently received renewed attention \cite{giddings,gia,hawking}, leading to firewalls \cite{firewall}, EPR-ER bridges \cite{epr}, and the prediction of (quantum) {\it hair} \cite{andi,gia2,steve}. In this short note, I~argue that there is yet another puzzle regarding black holes in general relativity which resembles some of the --alleged-- tunings in particle physics, although at the {\it classical} level. However, unlike say the cosmological constant problem, this tuning is associated with the {\it vanishing} of some of the --otherwise permitted-- (Wilson) coefficients in the classical effective field theory (EFT) describing black holes in long-wavelength gravitational backgrounds \cite{review,nrgrLH}. Namely, the black hole's `Love numbers,' namely the multipole moments induced by a tidal gravitational field, are zero in classical general relativity (in four spacetime dimensions) \cite{damourlove,poissonlove,smolkinlove,nhbeck,abhay}.\vskip 4pt

The direct observation of gravitational waves (GWs) by the LIGO scientific collaboration~\cite{ligodetect} is one of the greatest achievements of the century, thus far. Time and again scientists have~compared this feat to Galileo pointing his telescope to the sky, offering instead an {\it ear} to the cosmos. After the remarkable landmark of detection, GW science will soon turn into the study of the properties of the sources. While the strong coupling regime is naturally posed to become a laboratory to test gravitational dynamics, e.g. \cite{steve,lab1,lab2,luis,tokyo}, the inspiral phase of binary coalescences will also carry vital information. This is the regime where the value of the Wilson coefficients may teach us a great deal about compact objects.\vskip 4pt
\begin{wrapfigure}{r}{0.2\textwidth}
\vspace{-0.7cm}
\begin{center}
\includegraphics[width=0.2\textwidth]{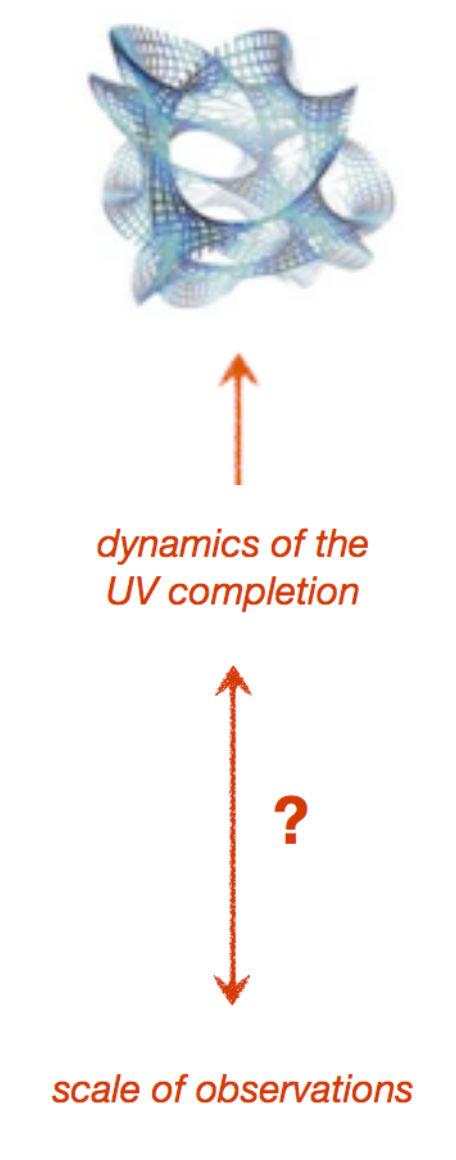}
\end{center}
\vspace{-1.3cm}
\end{wrapfigure}

Constraints on Wilson coefficients in an EFT framework have been used in the past to study fundamental properties of ultraviolet completions of gravitational theories, for instance the so called DGP model \cite{causal} (and also the pre-Higgs electroweak EFT \cite{ww}). This is even more relevant in early universe cosmology, since the universe resembles a (fixed-energy) {\it cosmological collider} \cite{cosmolo} where observations are bounded by the Hubble scale during inflation. Hence, measuring and/or putting bounds on these coefficients is a natural way to constrain putative quantum theories of gravity \cite{bmode,analytic}. In light of the recent direct detection of GWs with Advanced LIGO \cite{ligodetect,multim}, and by the same token, I~argue here that the vanishing of Love numbers in vacuum classical general relativity opens a unique venue to test the physics of spacetime, including phenomena such as the `axiverse' \cite{axiverse,axiverse2} and the very nature of black holes~\cite{steve,smatrix}, through precision GW~measurements. 
\vspace{-0.3cm}

\section*{Love is Tuned}
For over a hundred years physicists across disciplines have fallen {\it in love} with general relativity and black holes. However, this is unfortunately (or fortunately) not reciprocal, since both the black hole's electric- and magnetic-type Love numbers vanish, at the classical level\cite{damourlove,poissonlove,smolkinlove,nhbeck,abhay}. This result is somewhat puzzling, as I argue in what follows.\vskip 4pt In electrodynamics, applying an external electric field, $\vec{E}_{\rm ext}(\omega)$, induces a dipole moment (per unit of volume), $\vec{P}(\omega)$, in an object. The proportionality factor is the susceptibility, \beq \vec{P}(\omega) = \chi_e(\omega) \vec{E}_{\rm ext}(\omega),\eeq  as originally introduced by Lord Kelvin. A similar expression, with $\chi_e \to \chi_m$, $\vec P \to \vec M$ and $\vec E \to \vec B$, holds when applying a magnetic field \cite{jackson}. In the low-frequency limit, $\chi_e(\omega \to 0)$ accounts for the susceptibility response to a slowly changing (adiabatic) electric field. Next, there is an imaginary part proportional to $\omega$ which encodes the absorptive properties of the material, and so on. The susceptibility depends on the internal properties of the material. For instance, it modifies the index of refraction for light in a medium, $n(\omega)$. In~most cases --other than the vacuum-- we find $n(\omega \to 0) \neq 1$, which requires $\chi_{e(m)} (\omega \to 0) \neq 0$.\footnote{The real and imaginary parts of $\chi_{e(m)}(\omega)$ may be related through a Kramers-Kronig relation. Provided $\chi_{e(m)}(\omega) \to 0$ as $\omega \to \infty$, causality forces $\chi_{e(m)}(\omega \to 0)$ to be~non-zero. See \cite{analytic,fluid} for somewhat related discussions in the context of inflation and fluid dynamics. The possibility to set up sum rules in gravity was also discussed in \cite{iragrg,walter,review}.} 

\vskip 4pt

Something similar occurs in gravity, objects deform due to tidal forces induced by long-wavelength perturbations, and the equivalent of $\chi_{e(m)}$ in the zero-frequency limit are the electric- and magnetic-type Love numbers. In general, one can parameterize the response to gradients in the gravitational field in terms of multipole moments. Following the electromagnetic example, the response to an external field can be written as follows (at leading order in spatial derivatives) \cite{damourlove,poissonlove,dis1,dis2}
\beq
\label{responseret}
Q_E^{ij}(\omega) =  f_e(\omega)  E_{\rm ext}^{ij}(\omega)\,.
\eeq
%using the $SO(3)$ invariance of the black hole solution in a locally-flat co-moving frame $e^i_\mu$ (for non-rotating objects).%\footnote{This is reminiscent of the so called AdS/CFT correspondency, where isometries of the spacetime metric are mapped into global symmetries in the field theory side. The inclusion of spin is straightforward \cite{review}.} 
The $Q_E^{ij}$ is the induced (electric-type) quadrupole moment, and $E^{ij}$ is the (symmetric trace-free) electric component of the Weyl tensor projected onto a locally-flat co-moving frame, $e^i_\mu$. A similar expression holds for the magnetic-type quadrupole. In the Newtonian limit, with a static external gravitational potential, $\Phi_{\rm ext}(\vec x)$, we have
\beq
E^{ij}_{\rm ext}(\vec x) \propto \big(\partial_i \partial_j - 1/3\, \delta_{ij} \partial^2\big) \Phi_{\rm ext}(\vec x)\,.
\eeq
On purely dimensional grounds, for a (compact) object of mass $M$ and size $R$ we expect
\beq
\label{dimanal}
f_e(\omega \to 0) \sim M R^4 \sim \frac{R^5}{G_N} ,
\eeq
where we used $R \sim G_N M$. This dimensional analysis argument gives a correct estimate for the case of neutron stars, e.g. \cite{tanja}. However, the story takes a different turn for black~holes. \vskip 4pt

As in the electromagnetic case, in the low-frequency limit we can expand the function $f_e(\omega)$ in powers~of~$\omega r_s$, with $r_s$ the Schwarzschild radius,\footnote{The factor of $\mu$, and logarithms, are due to the presence of divergences in the point-particle limit. These are handled by dimensional regularization. The $\mu$-independence leads to a renormalization group trajectory \cite{review}.}
\beq
\label{eq:Jan2}
f_e(\omega) = f^{(0)}_e(\mu) + \beta^{(0)}_e \log (\omega^2/\mu^2)+ i f^{(1)}_e (r_s^{-1})\,\omega r_s+ r_s^2 \omega^2\big(f^{(2)}_e (\mu) + \beta^{(2)}_e \log(\omega^2/\mu^2)\big)  + \cdots \,.
\eeq
The imaginary part is related to absorptive properties of black holes. A~{\it matching} computation, comparing with the black hole absorption cross section in general relativity in the low-frequency limit, yields (see \cite{dis1,dis2,review}) \beq \frac{2G_N}{r_s^5} f^{(1)}_e(r_s^{-1}) = \frac{1}{45}\, \,.\eeq  On~the~other~hand, the deformation is encoded in the real part (which is invariant under $\omega \to -\omega$). The latter allows us to extract the black hole's electric-type Love~number,~$f_e^{(0)}(r_s^{-1})$, and similarly for the magnetic-type components.\vskip 4pt The (conservative part of the) response can be matched onto a point-particle effective action~\cite{review,nrgrLH},
\begin{align}
\label{Se2}
S_{\rm eff}  &=  \int d\tau \int d^4x~ \delta^4(x-x(\tau))\Big[ -m  + \frac{1}{2} C_E \, E_{\mu\nu}(x) E^{\mu\nu}(x) \\ &+ \frac{1}{2} C_{\ddot E}\, u^\alpha(\tau) \nabla_\alpha E_{\mu\nu}(x)\, u^\alpha(\tau) \nabla_\alpha E^{\mu\nu}(x)  +\cdots \Big]\, \nn.\end{align}
%\newpage
\begin{wrapfigure}{r}{0.24\textwidth}
\vspace{-0.4cm}
\begin{center}
\includegraphics[width=0.24\textwidth]{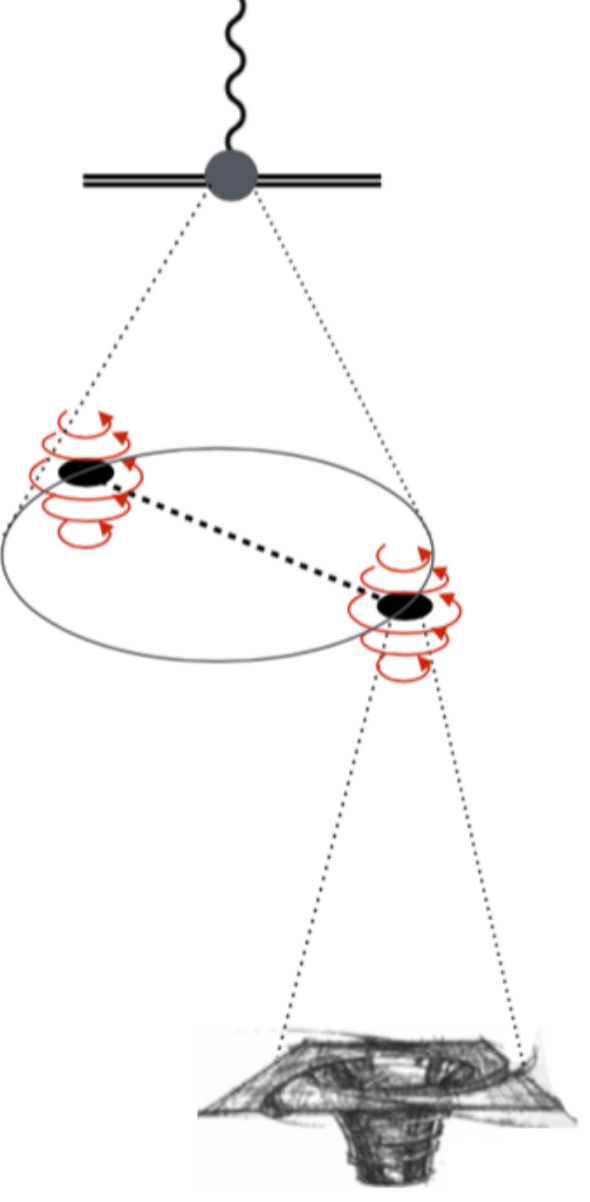}
\end{center}
\vspace{-1cm}
\end{wrapfigure}
%\hspace{-0.5cm} 
(The ellipses includes the magnetic-type couplings, $C_B$, etc.)\footnote{The absorptive part can also be incorporated in the EFT, as well as spin effects, see \cite{dis2,prd,prl,s1s2,s1s1,spinrad,spinrad2,review}.} The expression in \eqref{Se2} is the starting point of the EFT approach to gravitational dynamics \cite{review,nrgrLH,iragrg,eftgrg}. The extra parameters beyond `minimal coupling' are the Wilson coefficients. This effective action applies equally to black holes, neutron stars, or any other extended object in a gravitational long-wavelength background. The scaling of the extra terms in \eqref{Se2}, other than the mass, is ultimately connected with the effacement of internal structure in the two-body problem (for non-rotating objects). From the expressions in~\eqref{responseret} and~\eqref{eq:Jan2}, we notice that $C_E$ captures the information about the Love number.\vskip 4pt Due to the universality of the Wilson coefficients, we can obtain an estimate of the size of $C_E$, for instance, by considering the scattering of GWs off of~a~compact object.  The cross section in the EFT framework can be easily computed and, up to a numerical factor, we have
\beq
\sigma_{\rm eft} (\omega) = \cdots + G_N^2 C^2_{E(B)}\omega^8 +\cdots\, .
\eeq
Whereas in the fully relativistic computation for black holes we expect an expression of the form,
\beq
\sigma_{\rm gr} (\omega) = r_s^2 \,g(\omega,r_s)\,,
\eeq
with $g(\omega,r_s)$ some analytic function. (Non-analytic behavior cancels out in the matching.) 
Hence, expanding the cross section in powers of $\omega r_s \ll 1$ we find, again up to some dimensionless coefficient,
\beq
\sigma_{\rm gr} (\omega) \sim  r_s^2 \,(\cdots + (r_s\omega)^8+\cdots )\,.
\eeq
Comparing both expressions, and following the {\it naturalness} dogma of assuming all dimensionless quantities to be order one numbers, we arrive at~\cite{nrgrLH}
\beq
C_{E(B)} \sim  \frac{r_s^5}{G_N} \label{rs5}\,,
\eeq
which is compatible with the naive dimensional analysis expectation in \eqref{dimanal}. The value of $C_{E(B)}$ is universal, and can be used for instance, generalized to the study of neutron stars, to parameterize the relevance of the equation of state in the gravitational waveforms \cite{tanja}. Moreover, from the above scaling one concludes that finite size effects for compact objects first appear formally at ${\cal O}(v^{10})$ in the Post-Newtonian expansion, or 5PN order, for non-rotating bodies \cite{review,nrgrLH}. For black holes, however, an explicit calculation for the response yields~\cite{steinhoff}\footnote{~The function $f_e(\omega)$ in \eqref{responseret} follows from a retarded Green's function, and therefore it is analytic in the upper-half plane. The real part cannot be zero for all $\omega$ while having a non-zero imaginary part.}  \beq (f^{(0)}_e(\mu),\beta^{(0)}_e)~=~0\,~~ {\rm and}~~(f^{(2)}_e(r_s^{-1}),\beta^{(2)}_e)\neq 0\,.\eeq 
The vanishing of $f^{(0)}_e$ in vacuum~classical~general~relativity, together with the lack of logarithmic contribution at order $\omega^0$, implies: \beq \boxed{C^{\rm bh}_{E(B)}(\mu)=0}\eeq at {\it all} scales within the EFT realm. This turns out to be the case also for all of the induced electric- and magnetic-type multipole moments  in the $\omega \to 0$ limit \cite{poissonlove,damourlove}. 

\vskip 4pt The fact that Love numbers are zero for black holes clashes against some basic expectations. While the logarithms may not be present, we could still have power-law divergences arising in the EFT calculations requiring a `counter-term', $C_{E(B)}^{\rm ct}(\Lambda)$, proportional to an ultraviolet cutoff $\Lambda$. This counter-term may be needed at high Post-Newtonian order in the two-body problem \cite{nrgrLH,review}. Power-law divergences are set to zero in dimensional regularization mainly because all possible terms are already present in the effective action. Furthermore, we expect the coefficients to be determined by the relevant short-distance scale, in our case $r_s$. Since there is no apparent enhanced symmetry when $C_{E(B)}=0$, nothing prevents it from receiving (plausible large)~corrections. Hence, the fact that all of the Love numbers for black holes vanish --unprotected by symmetries-- implies a `fine tuning' from the EFT point of view \cite{iragrg,review}.\,\footnote{%~Something similar occurs in particle physics with the hierarchy between the electroweak and Planck scale, e.g. \cite{iratasi}. Since power-law divergences may be set to zero in dim. reg., one may claim there is no issue. Yet, one still cannot explain why $m_{\rm Higgs}$ is much {\it lighter} than a ultraviolet scale which in principle could be as high as the Planck mass. The problem worsens for the cosmological constant. 
Notice that, while I argue here the EFT mantra is pushing us toward non-vanishing Love, one could also turn the argument around, for example for the cosmological constant problem. In~full~general relativity one {\it explains} the vanishing of Love numbers for perturbed vacuum black hole solutions by an explicit calculation, at least at the classical level \cite{nhbeck,abhay}. Hence, another look at the problem may be that we have not yet found an ultraviolet completion which might elucidate the origin of the cosmological constant, `protected' by an equivalent {\it no-hair} theorem also in a quantum world.} 
\vspace{-0.3cm}
\section*{The  \textbf{\textit{Nature(ness)}}  of Spacetime} 
%\vskip 4pt 
The GW radiation produced by binary black holes (or neutron stars) during the inspiral phase, including finite size effects, can be tackled using the EFT framework \cite{nrgrLH,review}. The waveforms are computed as an expansion in the~velocity. The effect of the internal structure of the compact objects is encoded in higher derivative terms, as~in~\eqref{Se2}. The contribution from $C_E$ to the gravitational waveforms --the first such term for non-rotating bodies in the Post-Newtonian expansion-- encodes the imprint of degrees of freedom at the scale $r_s$ for an~extended~object. The Love number may be then obtained from the GW phase \cite{tanja}\footnote{To distinguish between a black hole and a neutron star binary system, we must either extract the total mass or identify an electromagnetic counterpart.}  
\beq
\delta \psi_{\rm Love} =  \frac{9}{16\eta M v^5} \sum_{i=1,2} c^i_E \left[ \left(12-11 \frac{m_i}{M}\right)v^{10} + \cdots\right]\,,\label{1}
\eeq
where $M$ is the total mass, $\eta \equiv m_1 m_2/M^2$ and~$c^i_E \equiv C^i_E/(m_i/M)$. Because of the analytic control over the inspiral regime  --in contrast to the intricate dynamics of the~merger-- the Love number has been suggested as a suitable candidate to test models of the equation of state for neutron~stars \cite{tanja}. At the same time, precision~GW measurements have also the potential to unveil details about black hole physics~in~a~cleaner~fashion. That is~the~case~since --in the jargon of particle physics-- there is no (classical)  `background' to be subtracted. In other words, a detection of a non-zero $C_E$ for black holes would point to {\it new} physics, due to the presence of (quantum) hair, e.g. \cite{steve,andi,gia2}, or `clouds' surrounding black holes, e.g. \cite{axiverse,axiverse2}.\footnote{In principle one can also test `modified gravity' models, e.g.~\cite{testgr1,testgr2}, although this is not the route advocated here.}\vskip 4pt Admittedly, we would expect quantum (renormalization) effects for large black holes ($r_s \gg l_{\rm P}$) to be highly suppressed, featuring an absurdly small --unobservable-- factor of $(\ell_{\rm P}/r_s)^n$ in~\eqref{rs5}, with $n\geq 1$. (This is the case perturbatively, if Planck-suppressed higher curvature terms are incorporated to the bulk gravitational effective action.)  In~fact, this is true for the one case we are most familiar with, namely Hawking radiation. In~the classical limit the temperature of a black hole vanishes, but in a quantum world we have ($k_B=1$)  
\beq 
T_{\rm bh} \simeq \frac{\hbar}{4\pi r_s}\,.
\eeq
This temperature is ridiculously small for astrophysical objects. For instance, for a solar mass black hole it is about $10^{-8}$ Kelvin. Hence, detecting Hawking radiation with astronomical observation is a daunting task. (In addition, one would have to subtract the cosmic microwave background.) However, precisely because of the --rather peculiar-- features of black hole physics needed to address the information paradox, including the alleged violations of locality at {\it long} distances \cite{giddings,steve}, it is not unthinkable one could wind up with a (non-perturbative) enhancement factor, 
\beq
C^{\rm bh}_E \stackrel{?}{=} N \left(\frac{\ell_{\rm P}}{r_s}\right)^n \frac{r_s^5}{G_N}\,,
\eeq
with large $N$.\footnote{Extended objects may prevent gravitational collapse through quantum effects, via `degeneracy pressure'. For~instance, for a neutron star the `enhancement factor' is given by $N_{\rm ns} \lesssim (M_{\rm P}/m_n)^3$, such that the radius scales like $r_{\rm ns} \lesssim \lambda_n\, N_{\rm ns}^{1/3} \sim 10\,$Km, with $\lambda_n = \hbar/m_n \sim 10^{-18}\,$Km, the~neutron's Compton wavelength. (Whereas for a compact  `boson star,' without self-interactions, we would have $r_{bs} \sim \hbar/m$, with $m$ the mass of the fundamental boson.) This demonstrates that (non-perturbative) macroscopical quantum effects are observable in the sky.} This would also be consistent with the ultraviolet sensitivity of the $C_E$ coefficient for black holes. Whether or not this turns out to be observable depends on the details of the ultimate theory, e.g. \cite{steve,andi,gia2}.\vskip 4pt Assuming the {\it natural} value for $C_E$, essentially dictated by dimensional analysis in~\eqref{rs5}, the expression in \eqref{1} starts formally at 5PN order (though it is enhanced for the case of neutron stars \cite{tanja}). Hence, this degree of accuracy becomes a well-motivated threshold to probe the {\it nature(ness)} of black holes.\footnote{In order to incorporate all possible effects at 5PN, the current state-of-the-art (approaching the 4PN order~\cite{4pn1,4pn2,nrgr4pn,tail}) must be pushed further yet an extra~notch. This entails also spin effects and the associated finite size terms \cite{review,prd,prl,s1s2,s1s1,spinrad,spinrad2,dis2,levi1,levi2,levi3}. Due to the complexity of the calculations, and in view of recent developments in gravity, e.g. \cite{jjtasi}, the search for a more efficient computational scheme within the EFT framework has also become an important goal in the field \cite{largeN,irashell,review}.} The vanishing of Love numbers in vacuum classical general relativity thus offers a fantastic opportunity to probe the very fabric of spacetime, in the advent of a new era of precision gravity.\newpage
\begin{center}
{\bf  \Large Acknowledgments}
\end{center}
I am grateful to Daniel Baumann, Dan Green, Sean Hartnoll, Eduardo Ponton, Ira Rothstein and Pedro Vieira for~useful conversations. This work was supported by the Simons Foundation and S\~ao Paulo~Research Foundation Young Investigator Awards, grants~2014/25212-3 and 2014/10748-5.

\end{document}